# Laboratory-based X-ray Absorption Spectroscopy on a Working Pouch Cell Battery at Industrially-Relevant Charging Rates

Evan P. Jahrman[1], Lisa A. Pellerin[1], Alexander S. Ditter[1], Liam R. Bradshaw[2], Timothy T. Fister[3], Bryant J. Polzin[3], Steven E. Trask[3], Alison R. Dunlop[3], and Gerald T. Seidler[1] (*)

[1]Physics Department, University of Washington, Seattle, WA  98195-1560
[2]Molecular Analysis Facility, University of Washington, Seattle, WA  98195-1560
[3]Chemical Sciences and Engineering Division, Argonne National Laboratory, Lemont, IL 60439

**ABSTRACT**

Li-ion battery (LIB) research has continuing importance for the entire range of applications from consumer products to vehicle electrification and grid stabilization.  In many cases, standard electrochemical methods only provide an overall voltage or specific capacity, giving an inadequate description of parallel redox processes or chemical gradients at the particle and pack level. X-ray absorption fine structure (XAFS) is frequently used to augment bulk electrochemical data, as it provides element-specific changes in oxidation state and local atomic structure.  Such microscopic descriptors are crucial for elucidating charge transfer and structural changes associated with bonding or site mixing, two key factors in evaluating state of charge and modes of cell failure. However, the impact of XAFS on LIB research has been significantly constrained by a logistical barrier: contemporary XAFS work is performed almost exclusively at synchrotron x-ray light sources, where beamtime is infrequent and experiment time-frames are limited.  Here we show that modern laboratory based XAFS can not only be applied to, e.g., characterization of *ex situ* LIB electrode materials, but can also be used for *operando* studies at industrially-relevant charging rates in a standard pouch cell preparation.  Such capability enables accelerated discovery of new materials and improved operation modes for LIBs.

 (*) seidler@uw.edu

1. Introduction

Lithium ion batteries (LIBs) serve diverse roles in the evolving modern energy landscape. This begins with small, consumer-level products, continues to vehicle electrification, and ends at several grid-level venues. While typical LIB applications require high energy density, a specification that is often achieved by developing cathode materials of high discharge capacity and high operating voltage,[1] recent industrial fast-charging ventures also demand improved power densities.[2] To meet these ends, there is strong evidence that the development of new battery chemistries and operation modes requires not just the usual bulk electrochemical characterization, but is instead strongly facilitated by studies using advanced x-ray spectroscopies. Specifically, x-ray absorption fine structure (XAFS) provides element-specific changes in local atomic structure (whether crystalline or disordered), bond lengths, and redox state.

It is useful to provide a brief synopsis of recent utilizations of XAFS in LIB research to showcase the utility of these measurements and the information attainable from them, especially in cases where multiple transition metals are included in the electrode chemistry. Of particular relevance to the present work, Mao et al.[3] paired XAFS with other x-ray techniques to investigate the fatigue of an Ni-rich Ni-Mn-Co (NMC) LIB material. Ni-rich NMC LIBs garner significant interest by extending the capacity accessible at reasonable voltage cutoffs.[4] The utility of these systems is, however, hindered by instabilities due to oxygen evolution during phase transitions and concomitant redox chemistry changes in the constituent transition metals.[4-6] These researchers found high-voltage charging did not result in obvious lattice changes or oxygen evolution, contrary to previous reports. Rather, increased disorder is found around the Ni sites and the oxidation state of the Ni atoms is highly heterogeneous, thus leading to high mechanical strain and micro-cracks. Note that this study was not an isolated effort in the literature. Indeed, the oxidation state of Ni-rich NMC materials has also been analyzed with XAFS by Tian et al.[7] and Bak et al.[5] Similarly, XAFS has been applied to the study of related materials. Aryal et al. examined all of the transition metals in a Li-rich Mn-Ni-Fe oxide cathode to reveal reduction of Mn and an irreversible loss of nearby oxygen atoms after cycling, while the Ni and Fe experienced little change in environment.[8] XAFS has also played a crucial role toward confirming charge transfer in analogous multivalent cathodes,[9,10] which are often difficult to analyze electrochemically due side reactions.[11] Furthermore, Kim et al. presented a wavelet transform analysis of Extended X-ray Absorption Fine Structure (EXAFS) measurements to reveal the Ni-O and Co-O distances change irreversibly and reversibly, respectively, during the first cycle of a NMC lithium ion battery, despite the redox reversibility of the Ni, Mn, and Co.[12]

The case of NMC batteries and the other mixed-metal systems, above, illustrate a common theme: The structural and speciation characterizations accessible by x-ray absorption spectroscopy are critical in developing new energy storage technologies, yet, until recently, this technique has been performed almost exclusively at synchrotron light sources and therefore subject to necessarily restrictive access models. This availability constraint is problematic for LIB research and development efforts. It limits the application of these techniques as valuable diagnostic tools in emerging battery materials requiring rapid feedback for further development, and it also hinders long-term studies, i.e., for degradation mechanisms, that require regular and extended access. Here, we show how the growing renaissance of lab-based, advanced x-ray spectroscopies can enable rapid and routine *operando* analyses of energy storage materials.

Some prior studies have used earlier generations of laboratory-based XAFS in energy storage research, this includes *ex situ* studies of Fe-substituted $LiCoO_2$ electrodes[13, 14] as well as *ex situ* studies of Fe-substituted $Li_2MnO_3$[15] and an *in situ* study of Fe-substituted $LiMn_2O_4$ charged at a rate of C/10.[16] However, *operando* x-ray analysis is widely believed to be a vital tool for monitoring element-specific oxidation in an inherently non-equilibrium process.[17] Fortunately, the field of laboratory-based x-ray absorption spectroscopy has seen rapid improvement in the quality of instruments based on both von Hamos[18-20] and Rowland[21-25] geometries. Indeed, these technologies are now being suggested as tools to address the development of nuclear fuels and disposal of subsequent waste,[26] as a platform for catalysis research,[27-29] and as a viable option for applying advanced x-ray spectroscopies to regulatory compliance testing.[30]

Here, making use of a high-powered laboratory XAFS user facility at the University of Washington (i.e., more than 1000 km from the nearest synchrotron x-ray light source), the x-ray absorption near edge structure (XANES) spectrum of Co, Mn, and Ni are measured at their respective K-edges in a working Ni-rich NMC battery during charging and discharging. All elements were measured at a rate of C/5 and we assess the redox behavior of each metal, but the Ni K-edge was also measured at faster rates, up to a maximum of 3C to establish the feasibility of laboratory-based *operando* studies at industrially relevant fast charge rates. Indeed, the x-ray intensities provided by the present spectrometer permit the pouch cell to be studied without modification, which the ensures the electrochemical behavior of the cell has not been perturbed.[31] Taken *en masse*, these results demonstrate a high potential for modern lab-based XAFS to impact LIB research and development.

2. **Experimental Details**

Single layer xx3450 pouch cell batteries were manufactured at the Cell Analysis, Modeling, and Prototyping (CAMP) Facility at Argonne National Laboratory. The anode used in these cells was Superior Graphite SCL1506T (graphite) and the cathode used was Toda NCM-04ST ($Li(Ni_{0.5}Mn_{0.3}Co_{0.2})O_2$ or NMC532). The composition of the anode electrode was 91.83 wt% graphite, 2 wt% Timcal C45 carbon black, 6 wt% Kureha 9300 (PVDF binder) and 0.17 wt% Oxalic acid. This electrode was coated onto a 10-µm thick copper foil. The electrode loading was 6.38 mg/cm$^2$ (coating only), electrode porosity was 37.4% and the electrode density was 1.36 g/cm$^3$ (no foil). The composition of the cathode electrode was 90 wt% NMC532, 5 wt% Timcal C45 carbon black and 5 wt% Solvay 5130 (PVDF binder). This electrode was coated onto a 20-µm thick aluminum foil. The electrode loading was 11.40 mg/cm$^2$ (coating only), electrode porosity was 33.1% and the electrode density was 2.71 g/cm$^3$ (no foil). Based upon anode and cathode loadings of the electrodes, the n:p ratio of the full cells is between 1.1 and 1.2, making this couple a balanced pairing.

After the electrodes were made, individual anode and cathode electrodes were punched to be made into the cells. The anode electrode size is ~32.4 mm x 46 mm and has an area of 14.9 cm$^2$. The cathode electrode size is ~31.3 mm x 45 mm and has an area of 14.1 cm$^2$. The anode is always slightly oversized to the cathode to prevent shorting between layers in a multilayer pouch cell. Tabs were ultrasonically welded to the electrodes and assembled in a wrap of Celgard 2320 separator (PP/PE/PP tri layer). A pouch was formed with pouch material from Youlchon Chemical (Cellpack-153PL) and the electrode assembly was placed in the pouch. A heat sealer was used to seal three sides of the cells. A quantity of (500 microliters) of 1.2 M $LiPF_6$ in (EC/EMC 3:7 wt%) was added to the pouch to serve as the electrolyte and solvent. Several shallow vacuum cycles were pulled to fully wet the cells before the final seal under vacuum was done. Formation cycles were performed at a C/10 rate, the cells were then degassed and resealed. The final cell dimensions were 48 mm wide by 75 mm tall (including tab length) and the cell exhibited a nominal capacity of 20 mAh.

During XANES measurements, all charging and discharging was performed with a LAND Battery Testing System, model CT2001A 5V1A, from Wuhan LAND Electronics Co., Ltd. C/5 rates were performed at the Ni, Co, and Mn K-edges via a nominal 4 mA current. For the Ni K-edge, cycles were also performed at rates of C/2.5, C, 2C, and 3C. Nominal charge rates were calculated assuming the ideal capacity of 20 mAh. In all XANES experiments, the cycle began with a rest period of approximately 10 minutes, a discharge at C/5 to 3.0 V, a 20 second pause, a charge at the appropriate rate to 4.1 V, a 20 second pause, discharge at the appropriate rate to 3.0 V, followed by another rest period of approximately 10 minutes. An applied voltage was not maintained during any rest periods. All measurements were performed on the same cell.

XANES measurements were performed at the University of Washington using the Clean Energy Institute X-ray Absorption Near Edge Structure (CEI-XANES) laboratory-based spectrometer. This general theory of operation of this instrument has been described in Seidler et al.,[24] and detailed descriptions of this particular implementation have also been presented.[23, 32] Briefly, measurements were performed using a (Siemens XFFAg4k) x-ray tube source operated with an accelerating potential of 20 kV and a tube current of 5-10 mA for $I_0$ and of 50 mA for $I_T$ of the pouch cell. The x-ray analyzers are various spherically-bent crystal analyzers with 1-m radius of curvature and 10-cm diameter (XRSTech). A small SDD (active area 25 $mm^2$, Amptek) was used to detect the x-rays in $I_0$ scans or after transmission through the pouch. Following our standard practice,[22, 24] $I_0$ scans are performed before or after the actual transmission-mode measurement; tube stability is sufficient to allow for this deviation from common methodology at synchrotron light sources.

All spectra were energy corrected by aligning the spectrum of an appropriate metal foil (Ni, Mn, or Co from EXAFS Materials, LLC.) to standards from `Hephaestus`.[33] All spectra were deadtime corrected. As the same location of the cell was probed for a set of measurements, the same pre- and post-edge lineshapes were applied to all spectra of the cell for a given element. For each K-edge, these lineshapes were acquired by fitting a spectrum collected on the battery under steady-state conditions and spanning several hundred eV. Spectra were then normalized according to the methods employed by `Athena` and `SIXPack`.[33, 34] Each XANES spectrum is then synchronized with the electrochemical characteristics (the voltage, capacity, and state of charge) measured at the start of its acquisition period. The Co K-edge was probed over several hundred eV. The Co and Mn K-edges were rebinned to provide satisfactory statistics using a floating summation of four or three subsequent scans, respectively. In order to extract the edge position for each spectrum, the data between approximately one quarter and three quarters of the white line intensity were first fit to a linear relation. The edge position was then chosen as the energy at which this function reached one half of the white line intensity. This is in general agreement with methods reported elsewhere.[35] In order to convert the XAFS spectra to photoelectron energy, each spectrum was shifted by its corresponding edge position. The position of the most distant scattering peak was then determined by assessing the mean of the photoelectron kinetic energy between 55 and 75 eV as weighted by the normalized absorption at each point.

### 3. Results and Discussion

The capacity and voltage values depicted as a function of time in Fig. 1 demonstrate healthy cycling performance for the selected battery. Several anticipated trends were observed, including a decreasing specific capacity at high charge rates coinciding with reduced Coulombic efficiency. In addition, the voltage profile exhibits sharp curves at the highest and lowest potentials, as is consistent with diffusion and polarization limitations. Furthermore, the plateau at intermediate potentials embodies significant structure as the thermodynamic availability of lithiation sites varies. Nonetheless, the charging potential was constrained to not exceed 4.1 V to avoid irreversible NiO formation due to oxygen evolution. As the present study encompassed several subsequent charging rates and edges, the present charging scheme was selected to avoid the degradation and irreversible phase transitions which have been previously reported.[36-38] The start of each XAS scan is also specified. The number of scans range between 38 during the C/5 charge and 10 in the discharge at 3C. In all cases, the number of scans acquired during a cycle was sufficient to provide insight into the electronic structure of the battery without substantial evolution of the sample between successive scans.

Beginning with Ni, we show the XANES spectra at the Ni K-edge in Fig. 2 as a function of charge state of the battery. For conciseness, we show only the C/5, C, and 3C charge rates. At each current, the Ni K-edge is observed to shift to higher photon energies as the battery is charged, and to lower photon energies as the battery is discharged. This behavior is consistent with the intercalation of $Li^+$ ions, as Ni is known to be the primary agent in charge compensation for similar materials.[39] By tracking the redox behavior of the Ni atoms, the state-of-charge of the battery may be directly assessed and, if desired, be constrained to a range which avoids the formation of undesirable Ni species. In the present instance, comparison to previously reported empirical standards[3, 40, 41] suggest that the present Ni oxidation state is between +2 and +3 in the discharged state and between +3 and +4 at full charge. Moreover, a strictly lateral displacement devoid of any isosbestic points suggests that the entirety of the Ni atoms evolve in concert rather than forming a linear combination, indicating a solid solution behavior rather than a mixed-phase behavior. This is significant for several reasons. Ni-rich NMC cathodes are liable to oxidation state heterogeneity across its constituent clusters of active material.[3] The lack of clear shoulders on the rising edge constrains the non-uniformity of the Ni oxidation state, while the lack of isosbestic points suggests that further redox activity does not strongly prefer Ni atoms of a given oxidation state. This provides further evidence for the healthy functioning of our cell as a dominant decay mechanism in Ni-rich NMC cathodes is deactivation by segregation of metal cations, irreversible structural changes, and isolation from the conductive notework.[42, 43] The linearity of the redox-dependent edge shift was further analyzed as in Fig. 3. A smooth and monotonic relationship between the edge position and stored charge was observed at each charge/discharge rate indicating the Ni is consistently redox-active throughout the charge cycle. This relationship is definitively linear, possessing a $R^2$ of at least 0.992 in all cases. Moreover, the roughly parallel nature of, e.g., the 3C and 1C charge rates coupled with the lack of points at diminished charge storage suggests that the reduced Coulombic efficiency is the result of incomplete conversion of Ni atoms back to their initial oxidation state.

Finally, the C/5 and C/2.5 rates were sufficiently slow to allow the Ni K-edge to be scanned over a more extended energy range. In these cases, a peak was observed due to the outgoing photoelectron scattering off nearby oxygen atoms. By shifting each spectrum by its edge position, the fine structure modulating $\mu$ can be plotted as a function of the energy above the K-edge, i.e., the kinetic energy of the photoelectron ($\Delta E$) for comparison, see Fig. 4. While the actual extended fine structure is difficult to extract over this limited energy range, the first peak is easily resolved. This peak shifts to higher energies as the state of charge increases, indicating a decrease in the Ni-O bond distance as is consistent with reports on similar systems.[44] The peak monotonically shifts to higher energies at the C/5 and C/2.5 rates, however it begins to plateau at higher states of charge for the faster charge rate. In addition, the photoelectron energy of the scattering peak does reaches higher values at the end of discharge than the start of the charge cycle. This is likely to be a result of the asymmetric condition caused by the applied voltage during use. Similar analyses relating the position of a scattering peak to a bond length have been pursued by other authors, especially in the field of actinide chemistry.[45, 46]

In addition to the electronic structure of Ni, the Mn and Co atoms were probed via *operando* XANES measurements at a C/5 cycle rate. The electrochemical measurements for both experiments are presented in Fig. 5. Spectra for Co were collected over several hundred eV and the most significant evolution was observed in the immediate vicinity of the K-edge, which is reported in Fig. 6. Here, two isosbestic point are observed on the rising edge of the Mn XANES spectra and one on the rising edge of the Co XANES spectra. For the Mn K-edge, the order of each spectrum between the isosbestic points is reversed outside of the isosbestic points. This behavior is consistent with the involvement of some Mn atoms in a disproportionation reaction, as has been proposed in other Li-ion systems.[47-49] However, the total lateral deviation is fairly minimal and the spectra are dominated by the +4 oxidation in all cases. Therefore, the rising-edge behavior may be due to subtler effects, including smaller changes in covalency. In contrast, only modest differences are observed in the rising edge of the Co XANES spectra. A departure from this trend is observed past the isosbestic point, where the white line feature shows significant sensitivity to the charge state and, as with the Mn spectra, the white line feature is seen to shift to higher energies upon charge and to lower energies upon discharge. Yet in the case of lithiated cobalt oxides, the effect of oxidation on the edge position is known to be fairly muted. Here, the observed behavior is similar to previous observations of the Co K-edge in Li$_x$CoO$_2$, Li$_x$NiO$_2$ with partial substitution of Ni by Co, and NMC-333 systems which often support conflicting hypotheses.[50-54] Some authors interpret such spectra to mean the Co oxidation state does not vary upon charge, while others assert the changes are due to oxidation of the Co or formation of oxygen holes on the oxygen ions neighboring the cobalt ions depending on the cell's state of charge. Finally, the above trends in the Mn, Co, and Ni K-edges are in agreement with previous measurements conducted on NMC-333 by Bak et al.[51] and Petersburg et al.[54]

## 4. Conclusion

Variation in the oxidation state of Ni was observed at elevated charge rates of an NMC battery. This was observed via *operando* XANES measurements in a laboratory setting and supports the utility of modern laboratory-based instrumentation in energy storage research, even to the point of fast-charging studies. Indeed, present instrumentation can perform these analyses *in situ* without modification of the pouch cell and with time resolution relevant to fast charging applications. These instruments can be used to assess the state-of-charge and state-of-health in prototype systems that require faster feedback than conventionally available at synchrotron light sources. Similar measurements could be particularly informative for next generation energy storage materials, such as batteries with multivalent charge carriers[55] or anion redox mechanisms,[56] where the charge transfer site can be ambiguous. As a future direction, this laboratory-based paradigm can serve as a useful model for studies of degradation mechanisms which require frequent and long-term access.

## 5. Acknowledgments

EPJ and TTF were supported in part by the Joint Center for Energy Storage Research (JCESR), an Energy Innovation Hub funded by the U.S. Department of Energy, Office of Science, and Basic Energy Sciences. EPJ was also supported by a subcontract from the National Institute of Standards and Technology. Opinions, recommendations, findings, and conclusions presented in this manuscript and associated materials does not necessarily reflect the views or policies of NIST or the United States Government. Part of this work was conducted at the Molecular Analysis Facility, a National Nanotechnology Coordinated Infrastructure site at the University of Washington which is supported in part by the National Science Foundation (grant NNCI-1542101), the University of Washington, the Molecular Engineering & Sciences Institute, and the Clean Energy Institute. This material is based in part upon work supported by the State of Washington through the University of Washington Clean Energy Institute. For the cells produced by the CAMP Facility in this study, we gratefully acknowledge support from the U. S. Department of Energy (DOE), Office of Energy Efficiency and Renewable Energy, Vehicle Technologies Office. Argonne National Laboratory is operated for DOE Office of Science by UChicago Argonne, LLC, under contract number DE-AC02-06CH11357.

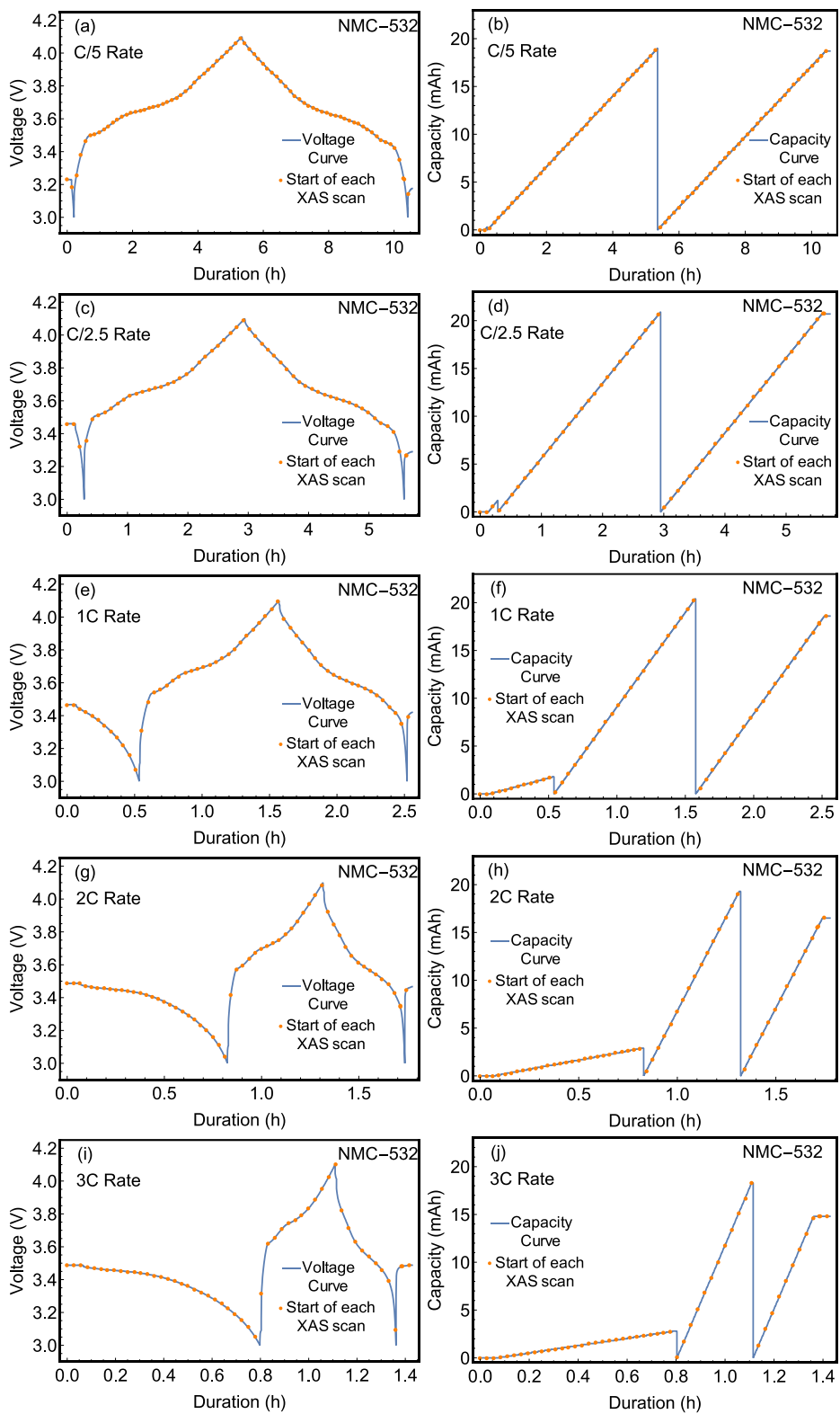

**Fig 1.** Measured voltage and capacity for the battery cell as a function of time from the start of the experiment. A series of dots designate the start of each XAS scan. Curves are shown for all cycles for which the cell was studied via XANES measurements at the Ni K-edge.

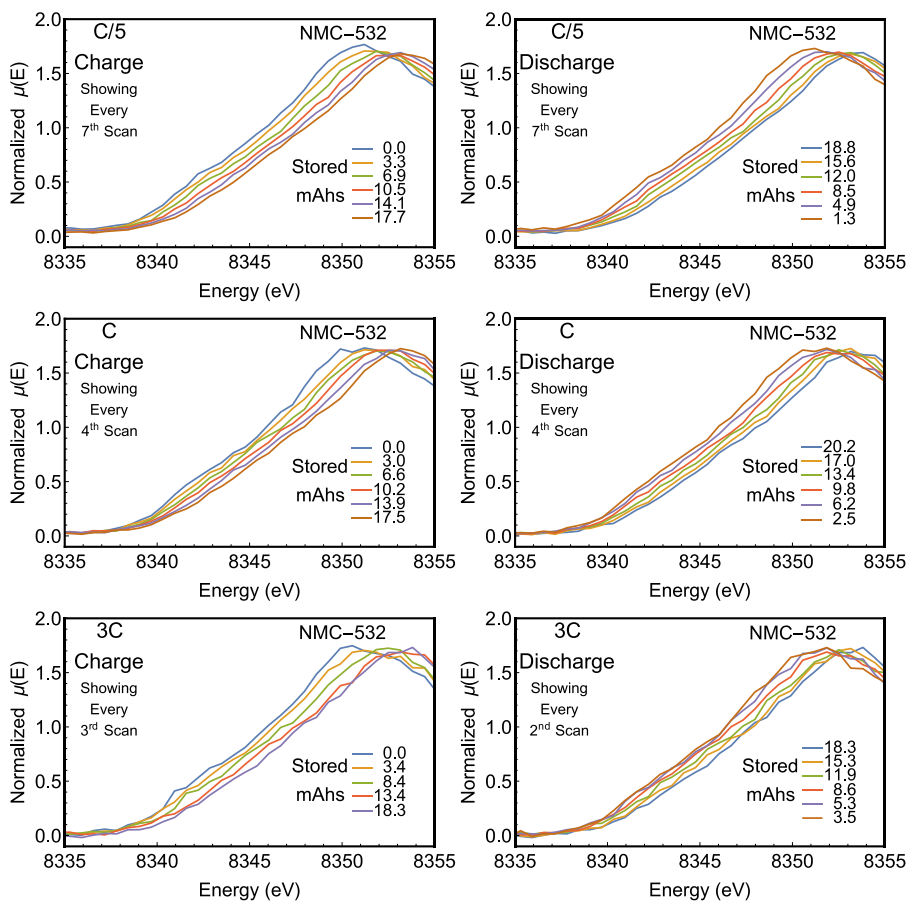

**Fig 2.** Ni K-edge XANES of the LIB at different degrees of lithiation. The state of charge at the start of each scan is provided in each figure. Only scans at specified intervals were provided for clarity. Spectra corresponding to a C/5 rate were truncated from scans extending over several hundred eV.

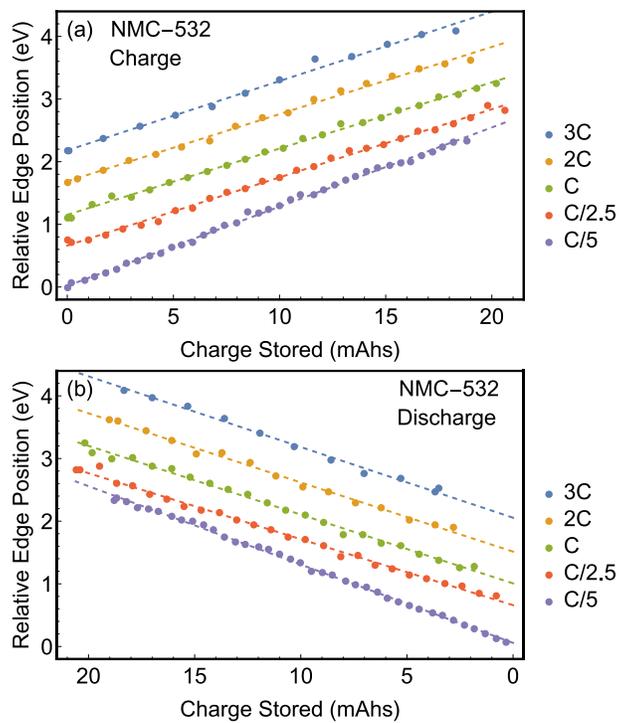

**Fig 3.** The Ni K-edge of each spectrum given as a function of state of charge at each cycle rate. Data taken faster than C/5 are offset successively by 0.5 eV.

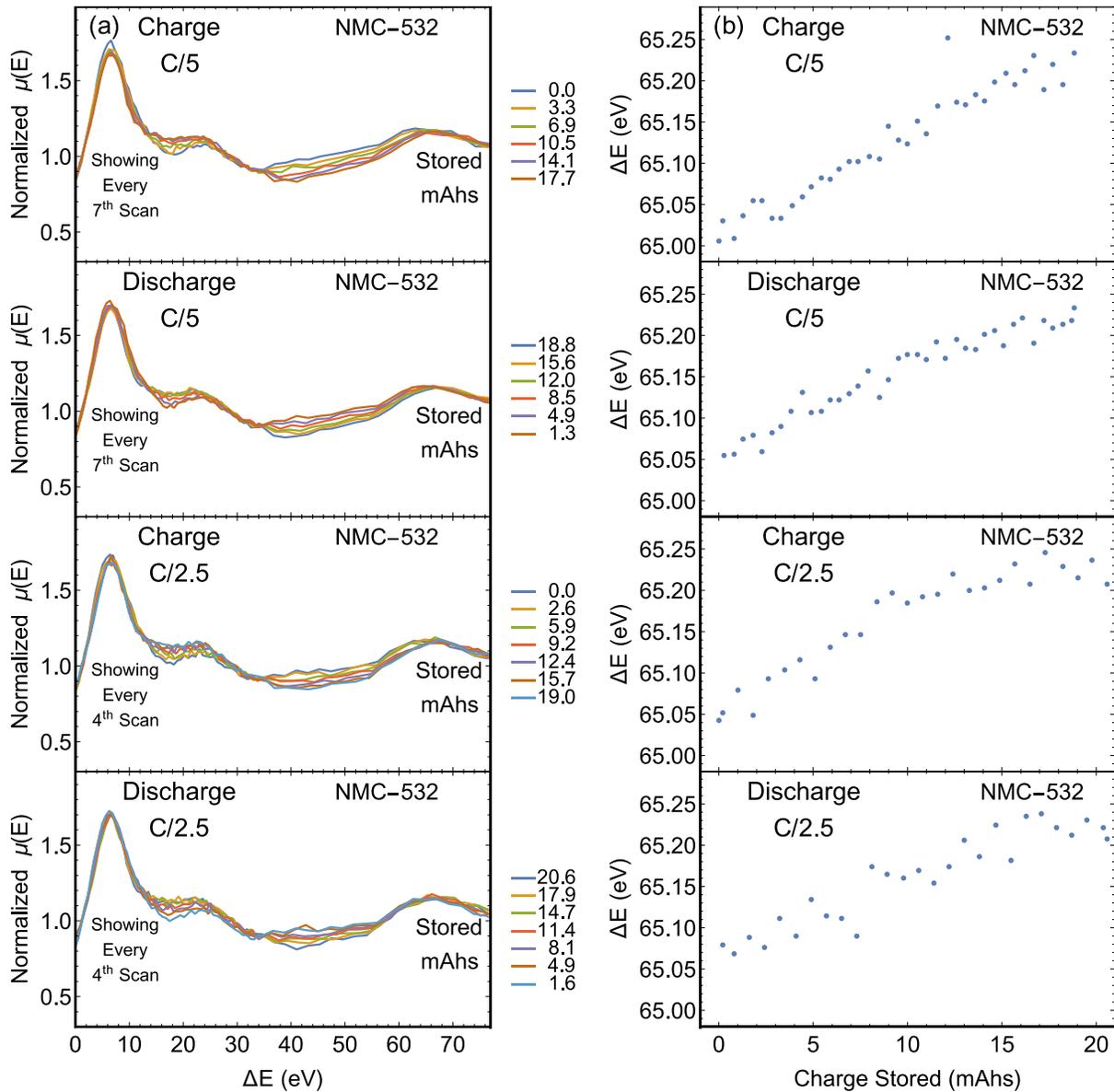

**Fig 4.** Ni K-edge XAFS of the LIB at different degrees of lithiation for the C/5 and C/2.5 cycle rates. The state of charge at the start of each scan is provided in each figure. The horizontal axis has been converted from energy to outgoing photoelectron momentum and the peak related to scattering from the neighboring oxygen atoms is around 4.3 Å$^{-1}$.

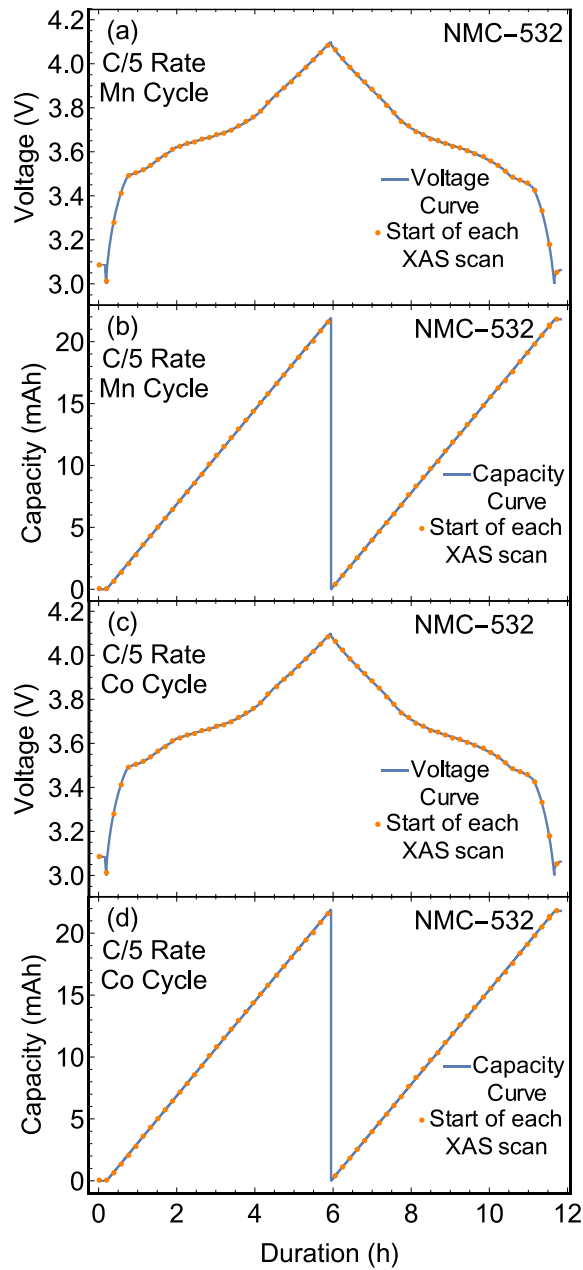

**Fig 5.** Measured voltage and capacity for the battery cell as a function of elapsed time. A series of dots designate the start of each XAS scan. Curves are shown for one full cycle during which XANES measurements were carried out at the Mn or Co K-edge.

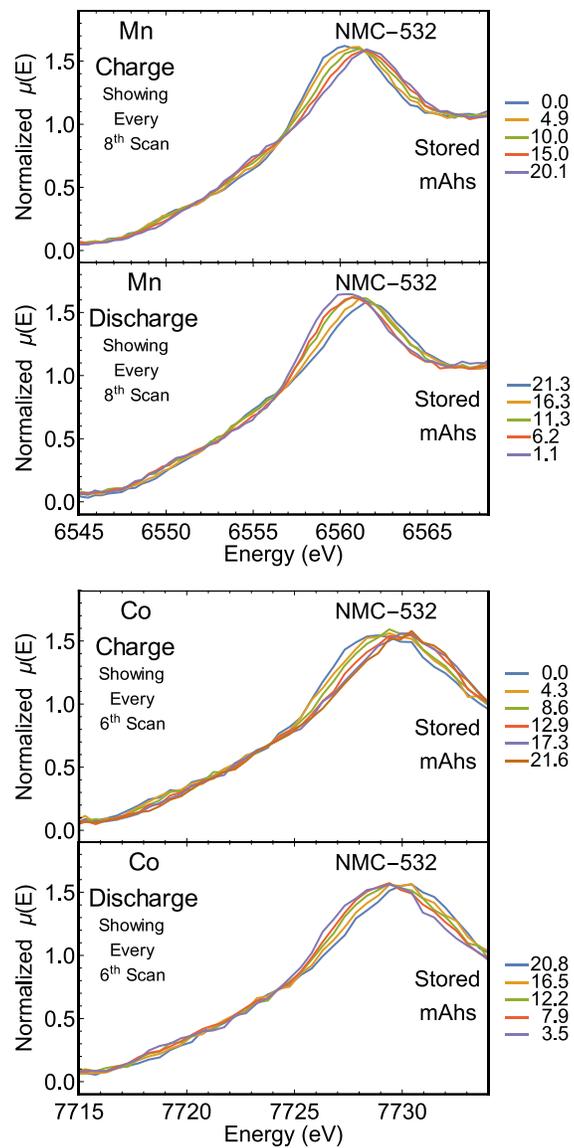

**Fig 6.** Mn and Co K-edge XANES of the LIB at different degrees of lithiation. The state of charge at the start of each scan is provided in each figure. Only scans at specified intervals were provided for clarity.

16